\documentclass[%
 reprint,
 amsmath,amssymb,
 aps,
]{revtex4-2}

\usepackage{graphicx}% Include figure files
\usepackage{dcolumn}% Align table columns on decimal point
\usepackage{bm}% bold math
\usepackage{hyperref}% add hypertext capabilities
\usepackage{threeparttable}
\usepackage{multirow}

\begin{document}

\preprint{APS/123-QED}

\title{Quantifying navigation complexity in transportation networks}% Force line breaks with \\

\author{Zhuojun Jiang}
\author{Lei Dong}%
 \email{arch.dongl@gmail.com}
\author{Lun Wu}
\author{Yu Liu}
 \email{liuyu@urban.pku.edu.cn}
\affiliation{%
 Institute of Remote Sensing and Geographical Information Systems, School of Earth and Space Sciences, Peking University, Beijing, China.
}%

\date{\today}% It is always \today, today,

\begin{abstract}
The complexity of navigation in cities has increased with the expansion of urban areas, creating challenging transportation problems that drive many studies on the navigability of networks. However, due to the lack of individual mobility data, large-scale empirical analysis of the wayfinder's real-world navigation is rare. Here, using 225 million subway trips from three major cities in China, we quantify navigation difficulty from an information perspective. Our results reveal that 1) people conserve a small number of repeatedly used routes, and 2) the navigation information in the subnetworks formed by those routes is much smaller than the theoretical value in the global network, suggesting that the decision cost for actual trips is significantly smaller than the theoretical upper limit found in previous studies. By modeling routing behaviors in growing networks, we show that while the global network becomes difficult to navigate, navigability can be improved in subnetworks. We further present a universal linear relationship between the empirical and theoretical search information, which allows the two metrics to predict each other. Our findings demonstrate how large-scale observations can quantify real-world navigation behaviors and aid in evaluating transportation planning.

\end{abstract}

\maketitle

%\tableofcontents

\section{\label{sec:level1}Introduction}

With the expansion of urban transportation networks, the efficiency and navigability of cities have attracted increasing attention \cite{gendreau_locating_1995,boguna_navigability_2009,de_domenico_navigability_2014,dong_population-weighted_2016,latora_efficient_2001}. In particular, the rapidly increasing number of edges (routes) in transportation networks has increased the complexity of navigation while making it easier for people to move around a city \cite{colak_understanding_2016,gallotti_anatomy_2015}. However, interestingly, people are not ``lost'' in complex transportation networks, which makes us wonder how individuals navigate during network evolution. The exploration of this question has important implications for understanding the correlation between the navigability of transportation networks and public travel behaviors.

Quantifying the routing costs in networks has offered novel insights into the navigation problems of transportation networks \cite{barberillo_navigation_2011,rosvall_networks_2005,rosvall_searchability_2005,2016Lost}, brain networks \cite{avena-koenigsberger_communication_2018,amico_centralized_2019,rajapandian_uncovering_2020}, social networks \cite{sneppen_hide-and-seek_2005}, wireless networks \cite{boushaba_node_2017}, and many other disciplines \cite{zanin_disorder_2008,yin_simplification_2020,perotti_smart_2012,cajueiro_optimal_2009}. Among these studies, the information approach proposed by Rosvall et al. \cite{rosvall_networks_2005,rosvall_searchability_2005} is an important starting point. By modeling navigation in a road network as a signal transmission process, Rosvall et al. developed a ``search information'' metric to quantify navigation difficulty \cite{rosvall_networks_2005,rosvall_searchability_2005}. Unlike route choice models, which are widely used in transportation engineering \cite{prashker_route_2004,prato_route_2009,kim_calibration_2020}, search information focuses more on measuring the complexity of the network (rather than modeling travelers' route choices and traffic assignment \cite{liu_transit_2010,raveau_behavioural_2014}). Specifically, the search information can be easily understood as the number of yes or no questions that a traveler has to answer when locating a route from a bird's-eye view of a map. However, previous studies on search information typically assumed that travelers have a global view of networks \cite{barberillo_navigation_2011,rosvall_networks_2005,rosvall_searchability_2005,2016Lost,zanin_disorder_2008}, ignoring the critical fact that travelers generally use information from only part of the network during their actual navigation \cite{muscoloni_navigability_2019,lee_exploring_2012,garling_introduction_2003}. More importantly, due to the lack of mobility data, previous studies often did not consider the actual traffic between network nodes, making it difficult to reflect the complexity of navigation in real networks.

\begin{figure*}[htbp]
	\centering
	\includegraphics[scale=0.5]{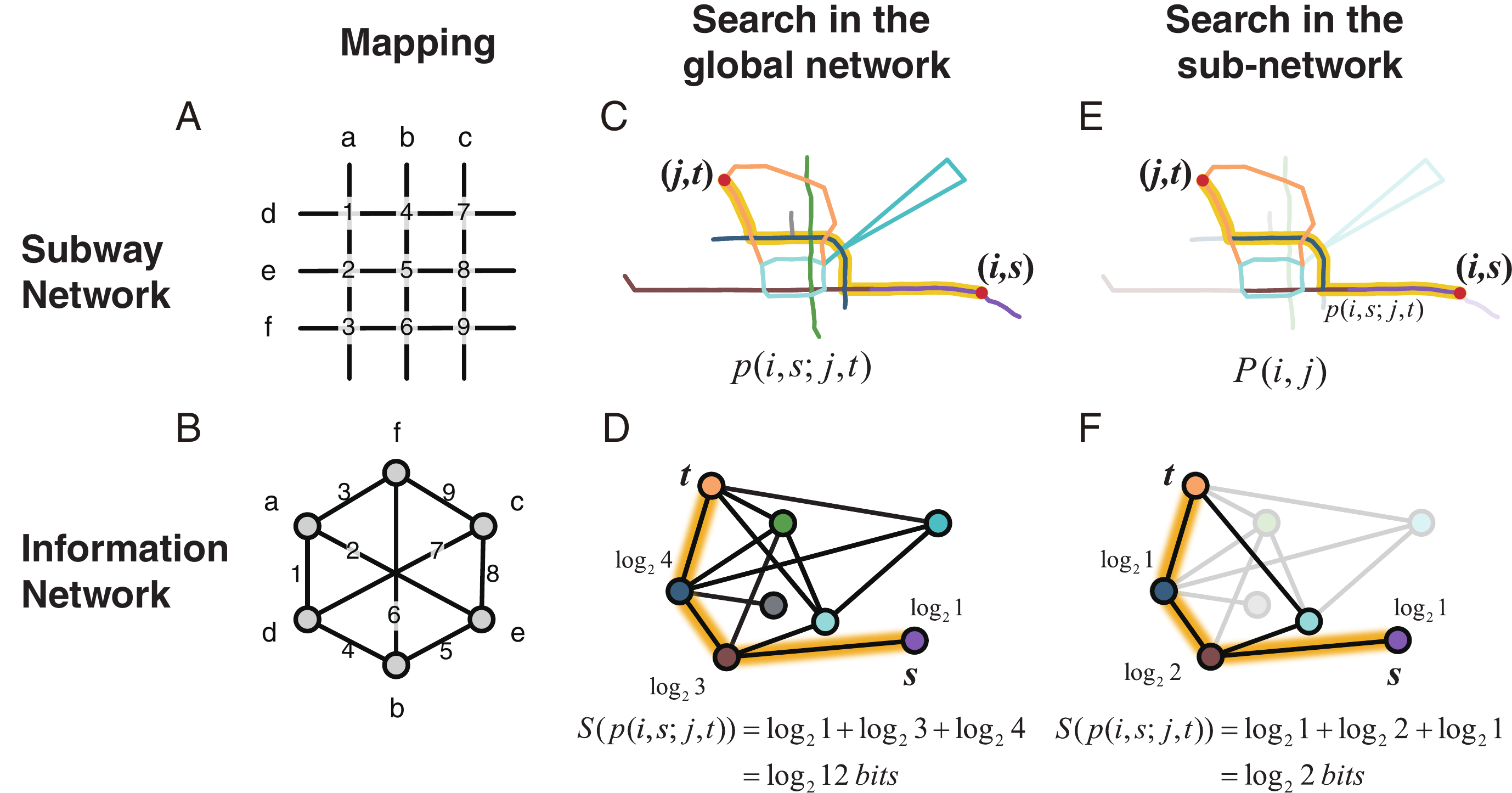}
    \caption{Search information. (A) A simplified schematic of a subway network. (B) The corresponding information network of panel (A). Nodes represent subway lines (a-f), and edges represent transfer stations (1-9). (C) Route $p(i,s;j,t)$ (highlighted in yellow) from station $i$ on line $s$ to station $j$ on line $t$. (D) Search in the global information network. One needs to answer $\log_2k$ yes or no questions to locate the next line or $\log_2(k-1)$ excluding the line that it has come from. (E) Search in the subnetwork. We delimit the subnetwork of actual trips by the stations and lines within the set $P(i,j)$ of route choices from $i$ to $j$. Subway lines without consideration are in faded colors. (F) The information network in solid colors is mapped from the subnetwork in panel (E). In contrast with the $3.6$ ($\log_{2} 12$) yes or no questions based on global search in panel (D), travelers only need to make one decision during their search in the subnetwork ($\log_{2} 2$), i.e., to decide whether to travel on the dark blue or light blue line.
    }\label{fig:network}
\end{figure*}

Here, using 225 million subway ridership records from Beijing, Shanghai and Shenzhen (three mega cities in China), we estimate the route for each origin-destination (OD) record and obtain the subnetwork formed by route choices to represent traveler networks for route planning. By mapping the subway network into an information network, we calculate the search information (the measure of navigability) along the path in the information network (Fig. \ref{fig:network}). According to ref. \cite{rosvall_networks_2005}, the $\log_2k$ bit is the minimal information necessary to locate the next move from $k$ options, where $k$ corresponds to the node degree in the information network [in Fig. \ref{fig:network}D, at each node except the origin node of the path, the incoming path is excluded from the $k$ options, resulting in an information value of $\log_2(k-1)$ bits]. Notably, the degree $k$ of the same node may be smaller in the subnetwork than in the global network, indicating that people are navigating with fewer line options (for transferring) in practice (Fig. \ref{fig:network}).

We observe that most people only make a small number of route choices during a trip, which implies that people use a small subnetwork when planning their routes. This directly results in lower search costs during actual travel than in theoretical studies \cite{rosvall_networks_2005,2016Lost,barberillo_navigation_2011}. By applying a simple route choice model to the historical subway networks of the three cities, we further investigate the impact of network growth on the decision information. We find that as the network grows, the navigation complexity of the subnetworks remains unchanged or even declines, while the theoretical global navigation complexity increases significantly. These differences suggest that the navigability of a network can be improved for actual travel behaviors, even if the global navigability is reduced during network growth. Moreover, we find a universal 3/4 linear relationship between empirical search information and theoretical search information, reflecting the hidden correlation between these two metrics in the complex network. This work bridges existing network navigability studies with traveler routing behaviors and has the potential to be used in evaluating urban transportation planning and understanding navigation in cities \cite{mckinlay_technology_2016,bongiorno_vector-based_2021,alessandretti_new_2021,coutrot_entropy_2022}.

\begin{figure*}[htbp]
	\centering
	\includegraphics[scale=0.5]{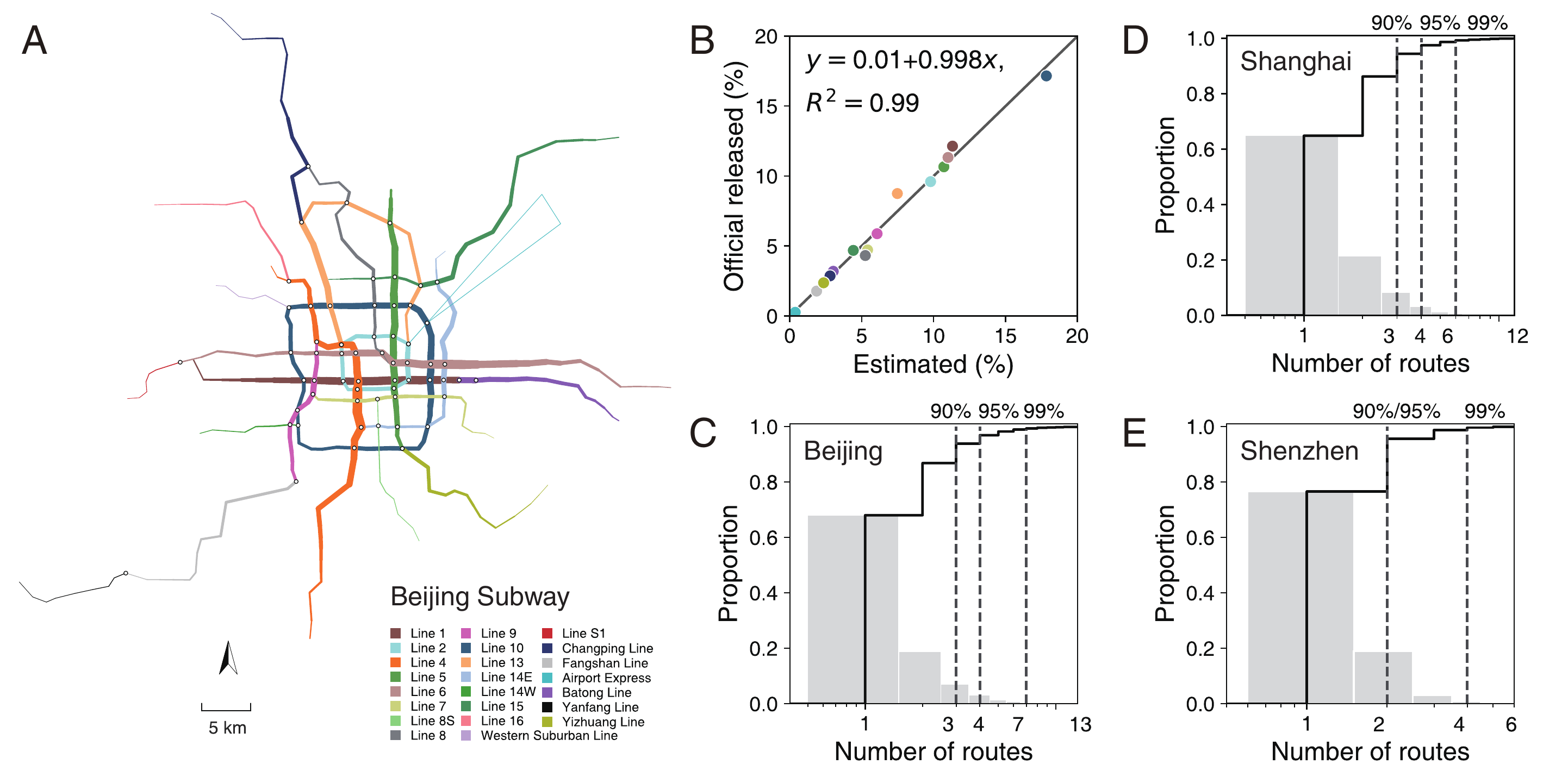}
    \caption{Route matching based on smart card records of subway systems.
    (A) The estimated flow map of the Beijing subway network.
    (B) Correlation between the estimated flow and the official released flow on 16 lines in the Beijing subway. Each point represents a subway line. The flow is normalized by the summation value and indicates proportions in the plot.
    (C-E) The number of routes followed by travelers between stations in the three studied cities. The distributions are displayed in a histogram (gray) and a cumulative distribution (black), respectively. The dashed lines mark the number of routes followed by travelers between 90\%/95\%/99\% of the station pairs.
    }\label{fig:flow}
\end{figure*}

\section{Results}

\subsection{The conserved number of routes}

We estimate actual traveler routes using smart card data, subway networks, and a travel survey dataset. Here we briefly introduce the data process; the detailed data description, the travel time estimation of a path and the discrete choice model can refer to the ``Materials and methods'' section. First, for a given station pair, the top 10\% and the bottom 10\% of smart card records are trimmed according to the travel time distribution to exclude abnormal data. Second, for each record, those paths in the subway network whose temporal distances are within 10 minutes from the travel time of the record are the candidate paths to be matched. Third, by adopting a discrete choice model considering travel time, distance, and the number of transfers, we estimate the probability of each candidate path being chosen. The parameters of the model are calibrated by the travel survey data (see the ``Materials and methods'' section). Finally, we match each record to the path with the largest probability of being chosen. To verify the matching results, we aggregate the flows by subway lines and calculate the correlation between the matched results and the official published number during the same period in Beijing (SI Appendix, Fig. S1) \cite{passvol}. The high goodness of fit ($R^2\approx0.99$) indicates the effectiveness of our matching method (Fig. \ref{fig:flow}B).

The matched results show that, despite the large number of possible routes between a station pair, most people only use a few routes (Figs. \ref{fig:flow}C-E). For example, in 95\% of the station pairs, people take no more than 4 routes in Beijing and Shanghai, and in Shenzhen, this ratio is 99\%. Among those limited number of routes, we further find that more than 90\% of people follow the simplest route to travel in the network (SI Appendix, Fig. S2), i.e., the route with the fewest transfers \cite{viana_simplicity_2013}. As more transfers usually mean higher time costs due to walking and waiting for the next transfer \cite{van_der_waard_relative_1988}, the simplest route can minimize transfer costs and tends to dominate people's choices.

\subsection{Information measures}
The adopted number of route choices observed in our data implies that actual travel occurs within subnetworks. It is possible that the information used for rider decisions consists of more than the subnetworks formed by the actual routes. However, since we can only observe traveled routes in our datasets, we assume that people's travel decisions are based on subnetworks formed by route choices in this paper. Here, we use the term ``route'' to denote people's route choices and the term ``path'' to denote the segments on the networks (the two terms are interchangeable).

First, we measure the navigation complexity in the information network via the information bits. Recall that a matched path $p(i,s;j,t)$ from station $i$ on line $s$ to station $j$ on line $t$ can be mapped to the information network (Fig. \ref{fig:network}F), where each node represents a subway line, and each edge represents the transfer station of the connected lines. According to ref. \cite{2016Lost}, the total amount of information $S$ needed to locate $p(i,s;j,t)$ in the information network is
\begin{equation}\label{eq:Su}
S(p(i,s;j,t)) = \log_2 k_s+\sum_{n \in p(i,s;j,t)}\log_2(k_n-1),
\end{equation}
where $k_s$ is the degree of node $s$. $k_n$ is the degree of node $n$ on path $p(i,s;j,t)$ (except $s$ and $t$), and one has to locate the next move from the $k_n-1$ options. $S$ measures the difficulty of locating a particular path, and a lower $S$ means that the network has better navigability between the station pair. When there are multiple matched paths between $i$ and $j$, we calculate the station-level empirical search information (ESI) by performing the flow-weighted average on all the matched paths $P(i,j)=\{p(i,s;j,t)\}$ connecting $i$ and $j$ (see the ``Materials and methods'' section). The value of ESI indicates how difficult it is for travelers to find their way from $i$ to $j$ during the actual trip.

Unlike the definition of search in the subnetwork formed by actual trips, the theoretical search information (TSI) defines the information required to find the simplest path in the global information network \cite{rosvall_networks_2005,2016Lost}, i.e., the path with the fewest nodes from $s$ to $t$. There may be multiple simplest paths between the stations; for simplicity, we use the fastest simplest path to define the station-level TSI according to ref. \cite{2016Lost}. The calculation for the TSI is the same as Eq. (\ref{eq:Su}), but the node degrees $k_s$ and $k_n$ are calculated in the information network transformed from the whole network.

\begin{figure*}[htbp]

        \centering
        \includegraphics[scale=0.5]{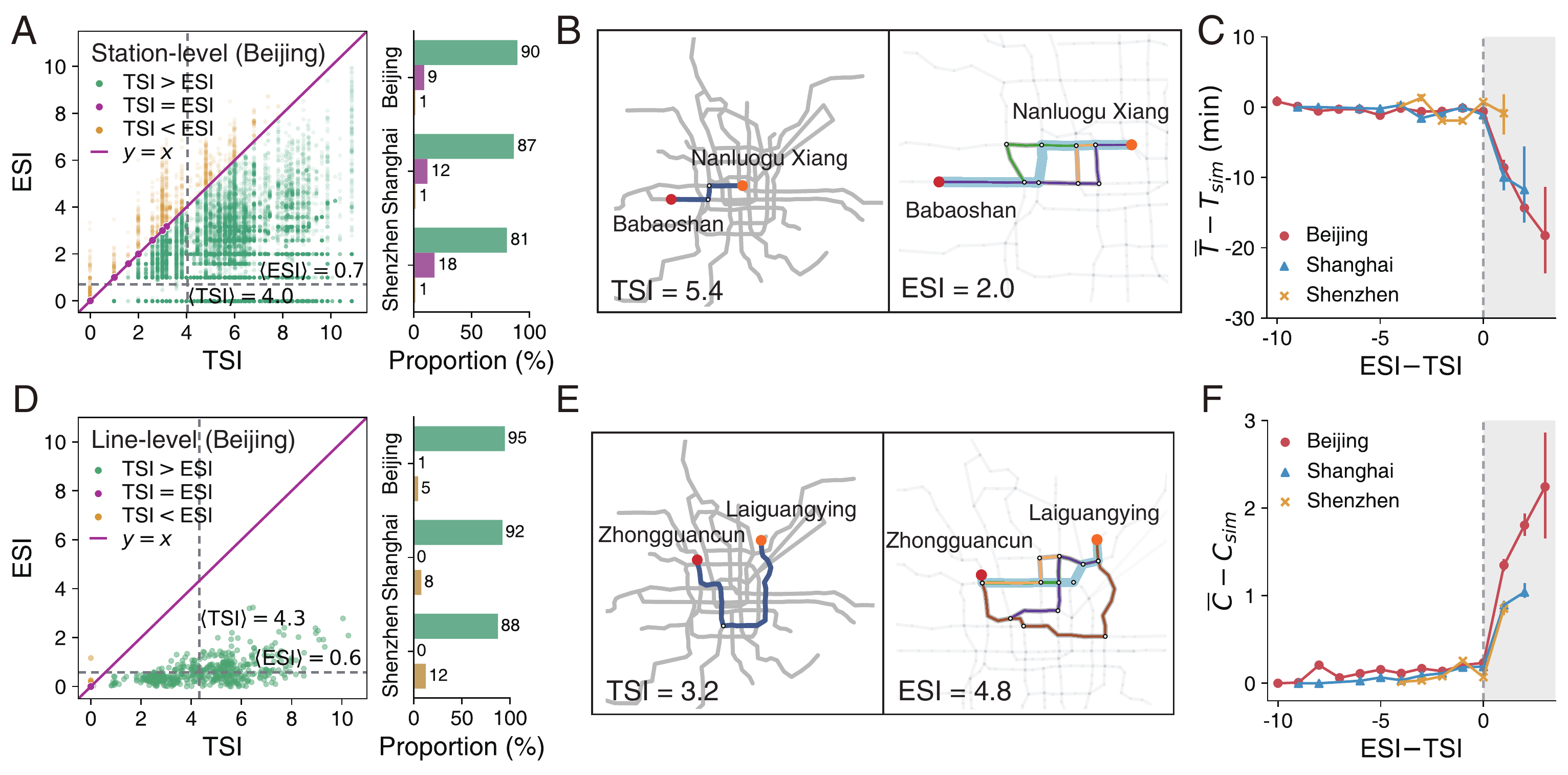}
        \caption{Station level and line level Search information.
        (A) The station-level ESI and TSI. In the left panel, each dot represents a station pair, and $\langle \cdot \rangle$ denotes the average over all pairs. In the right panel, the bars show the proportions of station pairs with TSI$>$ESI, TSI$=$ESI, and TSI$<$ESI.
        (B) An example of a station pair whose TSI$>$ESI. The subnetwork (right) formed by the 4 matched paths contains fewer lines and transfer stations than the global network (left), which makes ESI$<$TSI.
        (C) Time difference $\overline{T}-T_{sim}$ and information difference ESI$-$TSI at the station-level. For a station pair, $\overline{T}$ is the flow-weighted average of time for all the matched paths, and $T_{sim}$ is the travel time of the simplest path.
        (D) The line-level ESI and TSI. Similar to panel (A).
        (E) An example of a station pair whose TSI$<$ESI. Between this station pair, people choose the paths with more transfers ($\overline{C}=4.0$, $\overline{T}=61.6 \min$) in the subnetwork than the simplest path ($C_{sim}=1$, $T_{sim}=89.7\min$) in the global network, making the ESI exceed the TSI.
        (F) Transfer difference $\overline{C}-C_{sim}$ and information difference ESI$-$TSI at the station-level. The bars in panels (C, F) denote the 95\% CIs.
    \label{fig:Ss_distribution}
        }
\end{figure*}

Figure \ref{fig:Ss_distribution}A shows that the empirical search information values of most station pairs ($> 81\%$) are smaller than the theoretical global values. In the Beijing subway, the mean value of ESI and the mean value of TSI are approximately 0.7 bits and 4.0 bits, respectively, indicating that the decision cost for travelers to determine their routes is much smaller than what they theoretically need to spend. Similar results can also be found in Shanghai and Shenzhen (SI Appendix, Fig. S3). The main reason is that the subnetworks formed by route choices only include the lines and transfer stations that people have considered and adopted, and the number of these lines and transfer stations is far less than that of the whole network. This directly results in the node degrees of the subway lines in the subnetworks being smaller than those in the whole network when calculating the amount of information (Fig. \ref{fig:Ss_distribution}B presents an example).

Although the TSI is greater than the ESI between most stations, we find that a small number of station pairs in the three cities have an ESI equal to a TSI (Fig. \ref{fig:Ss_distribution}A). Most of these station pairs are located on the same subway lines without any transfers, and hence both empirical and theoretical decision costs are 0.

\begin{figure*}[htbp]
%   \begin{subfigure}[t]{0.5\textwidth}
        \centering
        \includegraphics[scale=0.5]{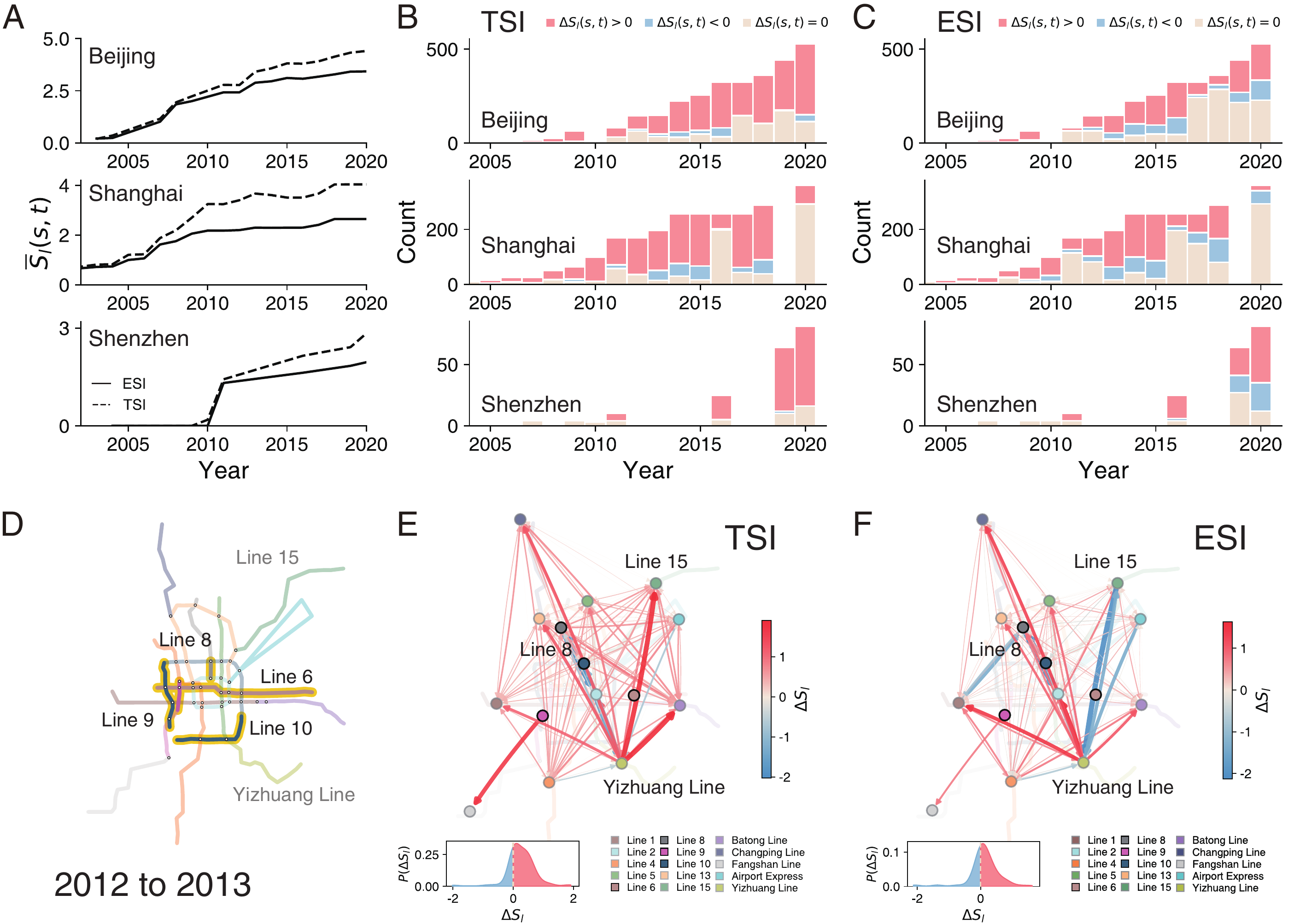}

    \caption{Line-level search information.
    (A) The development of $\overline{S}_l(s,t)$ for ESI and TSI.
    (B) Line-level TSI is decomposed into three categories. Each bin represents the number of line pairs in the subway network of the year. The colors distinguish the line pairs whose TSI values increase (red), decrease (blue) or remain unchanged (pale) due to the network changes in that year.
    (C) Similar to panel (B), the ESI is decomposed into three categories.
    (D) The network changes in the Beijing subway from 2012 to 2013. The newly opened segments/lines are highlighted in yellow.
    (E, F) The impact of network changes on navigability. The inter-line trips are drawn in the arcs that point from the start lines to the end lines. The colors indicate the differences in search information between 2012 and 2013. The extension of Line 10 connecting the Yizhuang Line makes navigation from the Yizhuang Line to other northern lines, such as Line 15 and Line 8, much easier (F), while it increases the difficulty of global navigation for most trips (E).
    \label{fig:S_timeline_diff}
    }
\end{figure*}

A counterintuitive finding is that there are still $1\%$ station pairs with ESI $>$ TSI (Fig. \ref{fig:Ss_distribution}A and SI Appendix, Fig. S3). This is mainly because people will trade higher navigation complexity for shorter times on some trips. Fig. \ref{fig:Ss_distribution}C shows that when ESI $>$ TSI (i.e., the higher navigation complexity), the travel time between stations is shorter than the simplest path time ($\overline{T}-T_{sim}<0$). Figure \ref{fig:Ss_distribution}E shows a typical example. From Zhongguancun to Laiguangying, most people choose shorter paths with $4$ transfers rather than the simplest path with $1$ transfer in the whole network. Since the search information is accumulated at each transfer along the path, the increased number of transfers (i.e., $\overline{C}-C_{sim}$) makes the ESI more likely to exceed the TSI (Fig. \ref{fig:Ss_distribution}F). Further analysis of station pairs with ESI $>$ TSI also verifies our assumption (SI Appendix, Fig. S4): the actual paths with more transfers are much shorter in travel time than the simplest path with fewer transfers.

In addition to the complexity of navigation between stations, another important dimension is the navigation between lines. In subway travel, the most important decision is how to best reach the destination through the choice of lines (rather than stations). Therefore, we average the amount of information on all the route choices that start on line $s$ and end on line $t$ to obtain the line-level search information $S_l$ between the line pair $s$ to $t$ (see the ``Materials and methods'' section). In Fig. \ref{fig:Ss_distribution}D and SI Appendix, Fig. S5, we again observe that more than 87\% of the line pairs have a lower ESI than TSI. The similar results between the station-level and line-level search information suggest that the small decision costs are universal across navigation dimensions in empirical navigability.

\subsection{Evolution of navigation complexity}
Subway systems in China have changed dramatically over the past two decades, giving us the opportunity to investigate the impact of network growth on navigation. To better understand the navigation complexity in historical networks without real-world trip data, we model navigation behaviors using the following steps: 1) generating a choice set, i.e., $m$ alternative paths by the repeated shortest path algorithm \cite{yen_finding_1971} and 2) making the route decision based on the discrete choice models used in matching smart card records with network paths (see the ``Materials and methods'' section). In the second step, we further assume that the path with the highest probability (among the $m$ paths) can represent the final route choice and use this path to measure the navigability in actual trips.

Specifying the size of choice set $m$ is challenging for route choice modeling, and it is usually based on laboratory settings due to the lack of empirical support \cite{bovy_modelling_2009,prato_modeling_2007}. Here, based on our findings (Figs. \ref{fig:flow}C-E), we use $m=13$ in Beijing, $m=12$ in Shanghai, and $m=6$ in Shenzhen to characterize the choice sets and infer the subnetworks that people use for route decisions. To validate the robustness of the results, we also conduct tests on different values of $m$ and obtain similar results (SI Appendix, Figs. S6-7).

Figure \ref{fig:S_timeline_diff}A shows that both ESI and TSI increase over time as the network grows larger, but the growth rate of ESI is substantially lower than that of TSI. To obtain a closer look at the changes, we classify the trips into three groups based on the changes in value from the search information. The results show that the search information for the vast majority of trips is increasing in the global network (Fig. \ref{fig:S_timeline_diff}B). In Beijing, for example, more than 70\% of the inter-line trips have an increased TSI. However, for search information in the subnetworks, we observe that most trips are not affected or are even easier to navigate during network growth (Fig. \ref{fig:S_timeline_diff}C). Especially since 2017, more than half of the inter-line trips have an unchanged or reduced ESI each year in Beijing and Shanghai, which reflects a great improvement in the actual navigability. When looking at the specific trips (Figs. \ref{fig:S_timeline_diff}D-F), we also find that adding subway lines can facilitate navigation for travelers, even if the difficulty of the theoretical global navigation increases.

\begin{figure*}[htbp]
%   \begin{subfigure}[t]{0.5\textwidth}
        \centering
        \includegraphics[scale=0.5]{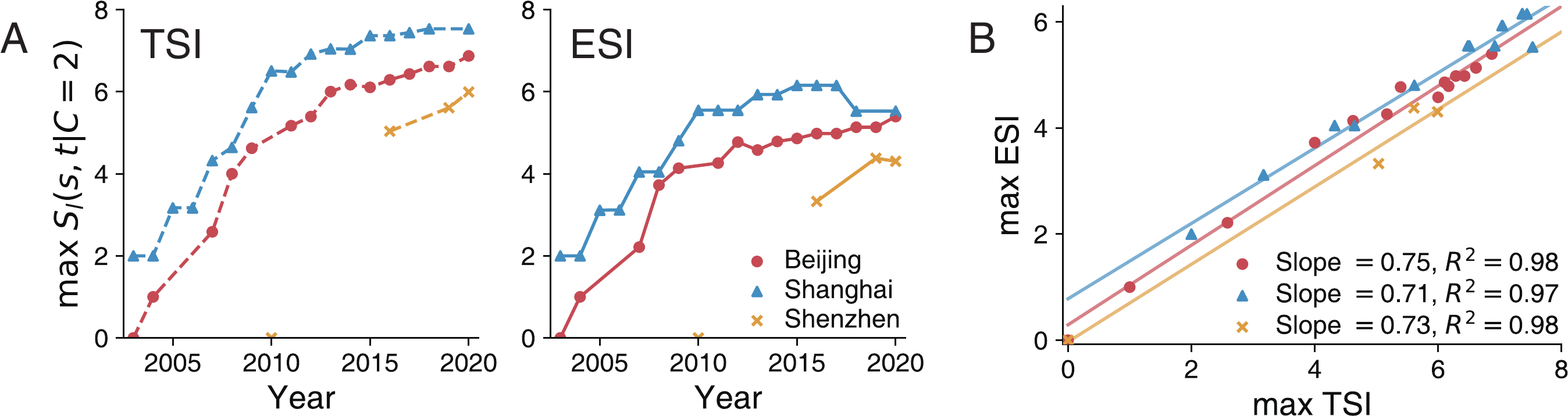}

    \caption{The relationship between line-level TSI and ESI.
    (A) The development of maximum TSI and maximum ESI for locating the fastest simplest paths with $C=2$ connections in the information network.
    (B) Maximum TSI versus maximum ESI. Each point represents the search information value of a year in panel (A). The maximum ESI is strongly correlated with the maximum TSI, and the slopes of the fitting lines are approximately 0.75.
    \label{fig:S_timeline}
    }
\end{figure*}

Figure \ref{fig:S_timeline_diff} focuses on the average value of the search information, while the maximum value is also essential for a transportation network since the maximum value determines the navigation obstacle. Similar to Fig. \ref{fig:S_timeline_diff}A, Figure \ref{fig:S_timeline}A shows that the maximum values of ESI and TSI increase as the network grows. In particular, the 8-bit upper limit of the TSI previously found in New York City \cite{2016Lost} can also be verified in Shanghai (Fig. \ref{fig:S_timeline}A). In addition to these specific values, we find a universal linear relationship with a slope of 3/4 between these two metrics despite the apparent complexity of network dynamics (Fig. \ref{fig:S_timeline}B). This finding expresses a very tight quantitative constraint on navigation behaviors and allows the empirical search information to be predicted by global search information. Given the 8-bit upper limit of the TSI and the 3/4 relationship, we estimate that the upper limit of the empirical search is approximately 6 bits, which is confirmed in the three cities (Fig. \ref{fig:S_timeline}A).

\begin{figure*}[htbp]
    	\centering
    	\includegraphics[scale=0.5]{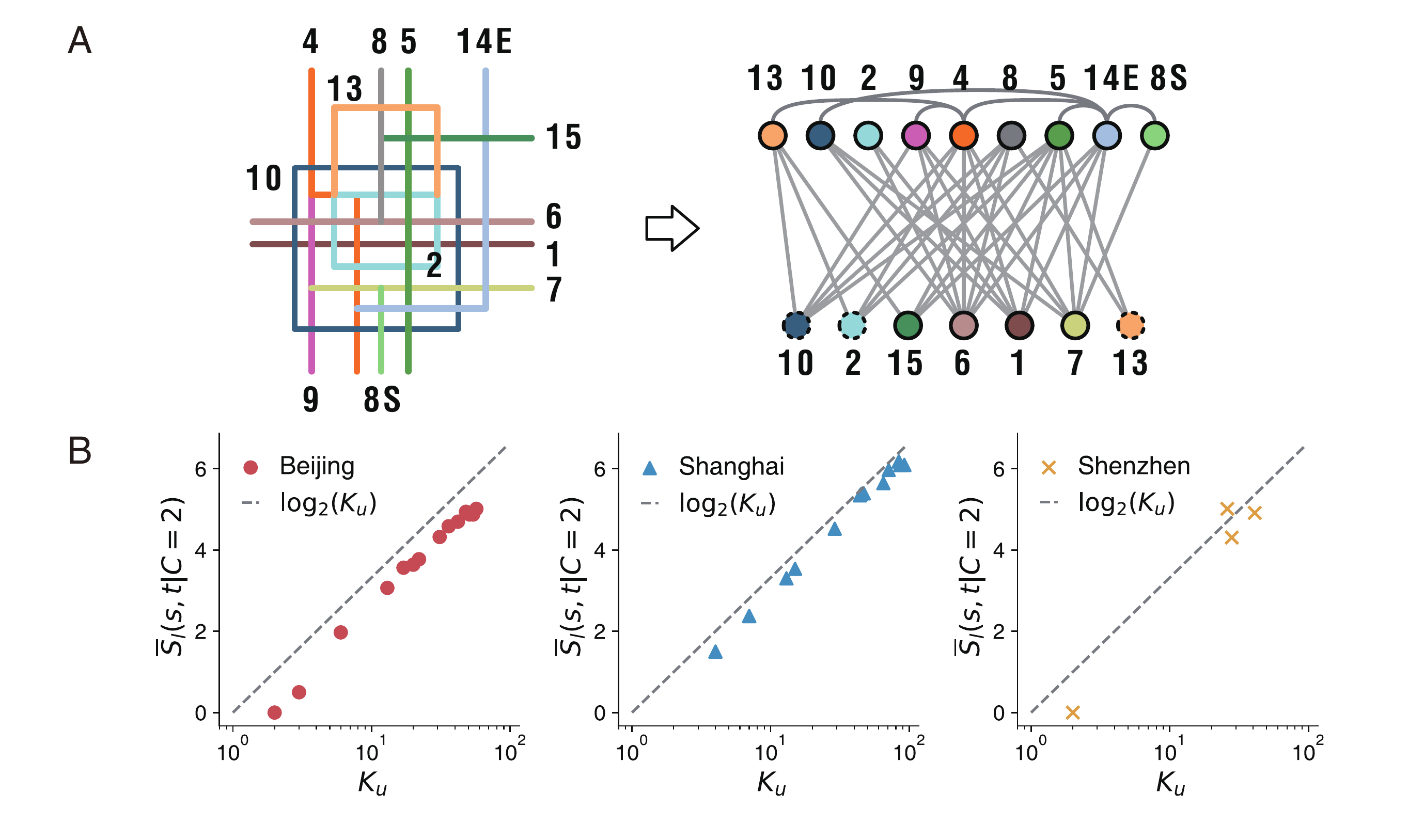}

    \caption{Line-level search information and the number of network connections. (A) Beijing subway network. Beijing subway lines have a grid-like structure (left), and the information network is similar to the bipartite map (right). As loop lines (Lines 2/10/13) act both as horizontal and vertical lines, we express them as two points (solid and dashed) with the same color in the information network.
    (B) Considering a path with 2 connections, the average amount of line-level search information needed to locate the simplest path is close to $S_l=\log_2{K_u}$, where $K_u$ is the number of edges in the information network or the number of connections in the subway network \cite{2016Lost}.
    \label{fig:lattice}
    }		
\end{figure*}

\section{Discussion}

In this work, we characterize traveler route choices from large-scale datasets and find that 1) people adopt a small number of repeatedly used routes and tend to prefer the simplest route; 2) the decision cost for wayfinding in the subnetworks is much smaller than that in a theoretical global search.
% 2) as the number of transfers or travel time increases, the decision cost on the subnetwork does not continue to grow as rapidly as global search.
In analyzing the search information required for actual navigation, we use the subnetworks formed by route choices. This may miss some additional routing information, as people may use more alternative routes for decision-making, even if those routes are never traveled (in the observed data). These unobserved routes would make search information values larger than the empirical results in this paper but would still be lower than the global theoretical values.

We discover a 6-bit information limit from dynamic networks, which coincides with the bound of the human working memory capacity of 2-6 items in cognitive sciences \cite{cowan_magical_2001,cowan_magical_2010,halford_separating_2007}. The 6-bit search information can also be linked to the number of connections in the network (which corresponds to the edges in the information network). According to ref.\cite{2016Lost}, the approximate relationship between search information $S_l$ and the number of connections $K_u$ of the information network is $S_l=\log_2{K_u}$ when considering paths with $C=2$ connections (note that this equation is derived assuming that the subway network is close to a regular lattice, as shown in Fig. \ref{fig:lattice}A; see \cite{2016Lost} for details). We also observe similar relationships between $\overline{S}_l(s,t|C=2)$ and the number of connections in the dynamic networks of the three cities (Fig. \ref{fig:lattice}B). Therefore, the 6-bit limit implies that people are disturbed by 64 ($2^6$) connections in the subnetwork during the search for the most complex trips.

Notably, both our 6-bit and the previous theoretical 8-bit limits are obtained when only the routes with 2 transfers are considered \cite{2016Lost}, and in fact, many trips require more than 2 (some require even 5) transfers to reach the destination. Therefore, the information limit summarized from the 2 transfer routes could be limited in representing the complexity of the network, and defining the navigation complexity at different scales (or transfer numbers) can provide a more comprehensive evaluation of the network structure \cite{guimera_optimal_2002}. Furthermore, identifying information limits may be critical, but reducing these travel complexities and designing a well-navigated transportation network are more profound.

Our work is helpful in evaluating transportation planning \cite{weckstrom_navigability_2021} and has implications for understanding human navigation in cities through large-scale datasets \cite{manley_shortest_2015,xu_collective_2017}. With the help of big data, quantifying real-world navigation behaviors may shed light on research in spatial cognition and psychology \cite{coutrot_entropy_2022,ramming_network_2001,epstein_cognitive_2017,chrastil_cognitive_2014}, which is usually based on experimental data from volunteer participants \cite{stangl_sources_2020,anggraini_neural_2018,liao_route_2017}. Our understanding of navigation complexity is still limited; especially due to the limitations of the data, we only study a subway network, which is a relatively simple public transportation network. Further research on multimodal networks (e.g., subways, buses and other modes of transportation) and spatial cognition mechanisms will provide a deeper understanding of urban navigation \cite{mckinlay_technology_2016,bongiorno_vector-based_2021,alessandretti_new_2021}.

\section*{Materials and methods}
\subsection*{Data}

The 225 million smart card records include 10.8 million users of the Beijing subway in May 2019, 10.3 million users of the Shanghai subway in April 2015, and 2.7 million in the Shenzhen subway in October 2017 (SI Appendix, Table S1). Each record consists of a card identification, the entry and exit timestamps, and names of the entry and exit stations and lines. For each record, the interval between the entry and exit timestamps is calculated as the travel time of that trip.

The subway network data include 15 snapshots of the Beijing subway from 2003 to 2020, 18 snapshots of Shanghai from 2000 to 2020, and 8 snapshots of Shenzhen from 2004 to 2020 (SI Appendix, Tables S2-4). We obtain the stations and lines from the official online subway maps. The coordinates of the stations and the travel time between stations are collected from Baidu Maps.

To calibrate the discrete choice model and validate the estimated travel time adopted in this paper, we collected 272 subway trips with known routes and durations through questionnaires in the three studied cities (SI Appendix, Figs. S8-9). Participants were asked to upload a screenshot of their subway ride record, including the entry and exit stations and times and the subway lines and transfer stations they took during the trip.

\subsection*{Travel time estimation and validation}

In a subway network, nodes represent stations, and edges represent subway lines connecting two consecutive stations. We set the in-vehicle time between a connected station pair as the temporal attribute of the edge and further assume that the travel time between the same adjacent stations between 2000 and 2020 is equal to the value in 2020 (not considering the effect of train speed increases).

To estimate the temporal distance of a specific path $p_{ij}$ from stations $i$ to $j$, we decompose the travel time into three parts: the in-vehicle time, transfer delay, and access/egress delay. We estimate each time separately: the in-vehicle time $T_{p_{ij}}^{veh}$ is obtained by the sum of the in-vehicle times on each edge; the transfer delay is assumed to be the product of the number of transfers $C_{p_{ij}}$ and the average transfer delay $T_{trans}$; the access/egress delay is assumed to be a constant value $T_{const}$. The temporal distance of $p_{ij}$ is
\begin{equation}\label{eq:2}
{T}_{p_{ij}}^{est} = T_{p_{ij}}^{veh} + C_{p_{ij}} \times T_{trans} + T_{const},
\end{equation}
where $T_{p_{ij}}^{veh}$ and $C_{p_{ij}}$ can be obtained from the subway network data. Although we assume that the average transfer delay $T_{trans}$ and the access/egress delay $T_{const}$ are constant at all stations, they are still difficult to observe. To solve this issue, we need to find some trips to be able to know their ${T}_{p_{ij}}^{est}$s and $C_{p_{ij}}$s and estimate $T_{trans}$ and $T_{const}$ as two regression coefficients. To do so, we assume that the trip with the least number of transfers among the $k$ shortest paths is the most time-efficient trip, whose travel time is the shortest (arrival time - departure time) derived from the card data. This trip is also known as the Pareto-optimal trip \cite{kujala_travel_2018}. Here, we use $k = 3$, and we also test this on different $k$s (SI Appendix, Fig. S10). ${T}_{p_{ij}}^{est}$ is estimated by the mean travel time of the time-efficient trips between stations $i$ and $j$. We then conduct the regression on the inputs of all station pairs to estimate the coefficients $T_{trans}$ and $T_{const}$ based on Eq. (\ref{eq:2}).

To validate the estimated travel time of the paths, we test Eq. (\ref{eq:2}) on the routes of the survey data. We calculate the expected time of each surveyed route in the network based on Eq. (\ref{eq:2}) and investigate the difference between this time and the actual travel time. SI Appendix, Fig. S11 shows that in all three cities, the time difference peaks at approximately $\Delta T=0$, and most trips (approximately 95.2\%) are in the range of $|\Delta T| \leq 10 \min$. The small gap verifies the effectiveness of the estimation (Eq. (\ref{eq:2})).

\subsection*{Discrete choice model}
To match the smart card records with routes on subway networks, we adopt the multinomial logit (MNL) model derived from utility theory, which is widely applied to traffic assignment \cite{janosikova_estimation_2014,kim_calibration_2020,prashker_route_2004}.

In the MNL model, the probability of choosing the path $p_{ij}$ from stations $i$ to $j$ in the choice set $P(i,j)$ is defined as

\begin{equation}\label{eq:logit}
Prob_{p_{ij}}=\frac{e^{V_{p_{ij}}}}{\Sigma_{P(i,j)}e^{V_{p_{ij}}}},
\end{equation}
where $V_{p_{ij}}$ is the deterministic utility function of $p_{ij}$ and is usually defined by a linear combination of factors affecting traveler route choices. We define the utility function by two explanatory variables as
\begin{equation}\label{eq:utility_V}
V_{p_{ij}} = \beta_0 \times T^{est}_{p_{ij}}+\beta_1 \times PI_{p_{ij}}^{trans},
\end{equation}
where $T^{est}_{p_{ij}}$ is the travel time of $p_{ij}$. $PI_{p_{ij}}^{trans}$ is the cumulative transfer penalty index \cite{kim_calibration_2020}, specified as $(1-e^{-C_{p_{ij}}})/{d_{ij}}$, where $C_{p_{ij}}$ is the number of transfers on $p_{ij}$ and $d_{ij}$ is the Euclidean distance between $i$ and $j$. This penalty index assumes that the impedance of transfers to people's route choice increases in a nonlinear form as the number of transfers increases, and the effect of this cumulative impedance is inversely proportional to the distance traveled \cite{kim_calibration_2020}. $\beta_0$ and $\beta_1$ are the parameters measuring the effects of $T^{est}_{p_{ij}}$ and $PI_{p_{ij}}^{trans}$.

To calibrate the MNL models for different cities, we estimate the parameters by the maximum likelihood method using the survey data. According to the estimation, the parameters set for Beijing ($\beta_0 = -0.0063, \beta_1 = -30.99$), Shanghai ($\beta_0 = -0.0023, \beta_1 = -127.7$) and Shenzhen ($\beta_0 = -0.0031, \beta_1 = -113.2$) are statistically significant (SI Appendix, Tables S5-7).

To validate the discrete choice models, we first calculate the matching accuracy on the survey data. Based on the probability of being chosen given by the MNL model, we match the path with the highest probability in the choice set to that record and check if this is the correct match. We finally obtain an overall correct rate of 92.1\% across the three cities, with small differences between each city (SI Appendix, Fig. S12). We then apply the models to the full set of smart card data and compare the aggregated flows on each line to the official published ridership numbers. The well-fitting regression also validates our models (Fig. \ref{fig:flow}B).

\subsection*{Search information}
We measure the difficulty of navigating a subway network based on search information \cite{rosvall_networks_2005} and quantify the navigation complexity at the station/line level by aggregating the information needed to follow a path.

\subsubsection*{Station-level and line-level search information}
The station-level search information is measured by averaging over all route choices $P(i,j)=\{p(i,s;j,t)\}$ from stations $i$ to $j$. Since locating each path $p(i,s;j,t)$ in $P(i,j)$ may require different amounts of information, we adopt a flow-weighted method for the aggregation
\begin{equation}\label{eq:Sij}
\begin{split}
&S_s(i,j) = \frac{\sum_{P(i,j)}S(p(i,s;j,t))\cdot  w(p(i,s;j,t))}{\sum_{P(i,j)}w(p(i,s;j,t))}.
\end{split}
\end{equation}
For a given station pair $i$ to $j$, ESI is calculated based on all the matched paths, and $w(p(i,s;j,t))$ is the weight value defined by the normalized flow on $p(i,s;j,t)$. For the TSI, the fastest simplest path is assumed to be the route choice between $i$ and $j$ according to ref. \cite{2016Lost}, and hence, only the single path is in $P(i,j)$, which simplifies Eq. (\ref{eq:Sij}) to $S_s(i,j)=S(p(i,s;j,t))$.

Similarly, we summarize the line-level search information by averaging over all route choices from line $s$ to line $t$
\begin{equation}\label{eq:Sst}
\begin{split}
&S_l(s,t) = \frac{\sum_{\{P(i,j)\}}S(p(i,s;j,t))\cdot  w(p(i,s;j,t))}{\sum_{\{P(i,j)\}}w(p(i,s;j,t))}.
\end{split}
\end{equation}
For the ESI, ${\{P(i,j)\}}$ contains all the matched paths whose origin station $i$ starts at line $s$ and destination station $j$ ends at line $t$. For the TSI, ${\{P(i,j)\}}$ contains all the fastest simplest paths between these stations.

\section*{Data and code availability}
The data and code used to reproduce the results are available at https://github.com/Jzjsnow/navi-complexity.

\bibliography{main}% Produces the bibliography via BibTeX.
% \begin{thebibliography}{99}

\end{document}